\documentclass[twocolumn,secnumarabic,amssymb, nobibnotes, aps, prd, superscriptaddress]{revtex4-2}
\usepackage{graphicx}
\usepackage{dcolumn}
\usepackage{bm}
\usepackage[T1]{fontenc}
\usepackage[utf8]{inputenc}
\usepackage{color}
\usepackage{array}
\usepackage{varwidth}
\usepackage{amsmath}
\usepackage{booktabs}
\usepackage{amssymb}
\usepackage{graphicx}
\usepackage{multirow}
\usepackage[english,main=english]{babel}
\usepackage{hyperref}
\usepackage{float}
\usepackage{makecell}
\usepackage{array}
\definecolor{darkblue}{RGB}{0,0,139}

\newcolumntype{M}[1]{>{\centering\arraybackslash}m{#1}}
\hypersetup{
    colorlinks=True,
    linkcolor=blue,
    urlcolor=blue,
    citecolor=blue
}

\begin{document}

\title{Symmetry-Based Microscopic Theory of the Unconventional Pairing Mechanism  in La$_5$Ni$_3$O$_{11}$ }%

\author{Guan-Hao Feng}
\email{fenggh@lingnan.edu.cn}
\affiliation{School of Physical Science and Technology, Lingnan Normal University, Zhanjiang, 524048, China}

\author{Jun Quan}
\affiliation{School of Physics, Changchun Normal University,  Changchun, 130032, China}
\affiliation{School of Physical Science and Technology, Lingnan Normal University, Zhanjiang, 524048, China}

\begin{abstract}
Recent experiments report high-temperature superconductivity in the hybrid nickelate $\mathrm{La}_5\mathrm{Ni}_3\mathrm{O}_{11}$, which is composed of alternating stacks of bilayer $\mathrm{La}_3\mathrm{Ni}_2\mathrm{O}_7$ and monolayer $\mathrm{La}_2\mathrm{NiO}_4$. However, the superconducting transition temperature $T_c \approx 64~\mathrm{K}$ for $\mathrm{La}_5\mathrm{Ni}_3\mathrm{O}_{11}$ is remarkably lower than the $80~\mathrm{K}$ observed for pressurized $\mathrm{La}_3\mathrm{Ni}_2\mathrm{O}_7$. Thus, an unified microscopic theory is required to address the difference in the pairing mechanisms between these systems. Here, we develop a phenomenological symmetry-based approach to systematically analyze the low-energy physics in $\mathrm{La}_5\mathrm{Ni}_3\mathrm{O}_{11}$, which is obtained by a charge self-consistent density functional theory plus dynamical mean-field theory method. We show that the superconductivity in $\mathrm{La}_5\mathrm{Ni}_3\mathrm{O}_{11}$ exhibits a two-gap nature, consisting of a leading interlayer pairing between the $d_{z^2}$ orbitals and a subleading intralayer pairing between the $d_{x^2-y^2}$ orbitals. The reduction of $T_c$ can be attributed to the diminished contribution of the interlayer pairing, as reflected by the hopping parameter ratio $|t_{\perp}^z/t_{\parallel}^{x}|$. Base on this unified picture, we discuss the possible pairing mechanism and the role of $\gamma$ pocket for the superconductivity in the bilayer NiO$_2$ planes of nickelate superconductors. 
\end{abstract}
\maketitle
\emph{Introduction.}---%
Superconductivity in pressurized bilayer (BL) La$_3$Ni$_2$O$_7$ has recently attracted a great deal of interest owing to the high superconducting transition temperature ($T_c$)~\cite{sun_signatures_2023,wang_normal_2024}. Other Ruddlesden-Popper nickelates, such as monolayer (ML) La$_2$NiO$_4$~\cite{PhysRevB.75.012414, PhysRevB.80.144523}, trilayer (TL) La$_4$Ni$_3$O$_{10}$~\cite{Li_2024, liStructuralTransitionElectric2024, zhuSuperconductivityPressurizedTrilayer2024}, hybrid ML-BL La$_5$Ni$_3$O$_{11}$~\cite{PhysRevMaterials.8.053401, nieSuperconductivityElectronicStructures2026, shiPressureInducedSuperconductivity2025, chen2026pairingmechanismsuperconductivity1313}, ML-TL La$_3$Ni$_2$O$_{7}$~\cite{nieSuperconductivityElectronicStructures2026, chenPolymorphismRuddlesdenPopper2024, PhysRevLett.133.146002, PhysRevB.110.155145},  and BL-TL La$_7$Ni$_5$O$_{17}$~\cite{nieSuperconductivityElectronicStructures2026}, have also attracted significant research interest. However, the behavior of $T_c$ varies significantly across these nickelate superconductors. Specifically, the optimal $T_c$ for ML-BL La$_5$Ni$_3$O$_{11}$ is $64$~K achieved at approximately $21$~GPa~\cite{shiPressureInducedSuperconductivity2025}, whereas the optimal $T_c$ for BL La$_3$Ni$_2$O$_7$ is $80$~K achieved at approximately $14$~GPa~\cite{nwaf220, Hou_2023, zhangHightemperatureSuperconductivityZero2024, PhysRevX.14.011040, zhang_effects_2024}. Theoretically, the low-energy physics of these systems is dominated by the Ni-$e_g$ orbitals~\cite{luo_bilayer_2023, PhysRevB.108.125105, PhysRevB.111.174506, lechermann_electronic_2023, PhysRevB.108.L180510, PhysRevB.109.L081105, yang_possible_2023, sakakibara_possible_2024, PhysRevLett.132.126503, PhysRevMaterials.8.044801, PhysRevB.108.165141, geislerStructuralTransitionsOctahedral2024, PhysRevB.109.144511}. This raises an open question---what leads to the reduction of $T_c$?

A recent random-phase approximation (RPA) calculation showed that the $d$-wave pairing in the SL subsystem of La$_5$Ni$_3$O$_{11}$ is the leading instability~\cite{1mr2-s6s8}. However, the dynamical mean-field theory (DMFT) have revealed that the SL subsystem is in Mott-insulating state and thus the superconductivity should arise from the BL subsystem~\cite{k83c-qgx1, leonov2026electronicstructurequasiparticlerenormalizations, 1412-nfzm, zhangPairingMechanismSuperconductivity2026b}. Moreover, the interlayer hopping $t^z_{\perp}$ in the BL subsystem between the $\mathrm{Ni-}3d_{z^2}$ orbitals may play an important role in $T_c$, as this hopping is widely regarded as a primary driver of interlayer electronic correlations~\cite{PhysRevLett.131.236002, PhysRevLett.133.096002, luo_bilayer_2023, PhysRevB.111.174506, PhysRevLett.134.076001, PhysRevLett.132.106002, zhang_structural_2024, qiu2025pairingsymmetrysuperconductivityla3ni2o7, PhysRevLett.131.236002, xia_sensitive_2025, yang_possible_2023, zhang_structural_2024, PhysRevB.108.174511, PhysRevB.108.214522, PhysRevB.108.L140504, PhysRevLett.132.146002, PhysRevB.109.165154, wuSuperexchangeChargeTransfer2024}. In this respect, a unified theoretical framework encompassing both ML-BL $\mathrm{La}_5\mathrm{Ni}_3\mathrm{O}_{11}$ and BL $\mathrm{La}_3\mathrm{Ni}_2\mathrm{O}_7$ is essential to uncover their microscopic pairing mechanisms. Inspired by the recent experimental observation that the superconducting transition in pressurized-bulk La$_3$Ni$_2$O$_7$ is accompanied by a structural phase transition~\cite{nwaf220}, the emergence of unconventional superconductivity may be closely related to symmetry. Notably, a shared feature of these two systems is their layer group symmetry, $p4/mmm$, which enables us to establish an unified symmetry-based approach to systematically investigate the microscopic pairing mechanism. 

In this Letter, we employ a charge self-consistent density functional theory (DFT)~\cite{QEref1, QEref2, PhysRevB.88.085117} plus DMFT method~\cite{Merkel2022, Seth2016274, triqs_ctqmc_solver_werner1, triqs_ctqmc_solver_werner2, triqs_ctqmc_solver_legendre, lewin_thesis, PARCOLLET2015398} to investigate the low-energy physics of ML-BL $\mathrm{La}_5\mathrm{Ni}_3\mathrm{O}_{11}$ subjected to strong local electronic correlations. We demonstrate that the ML subsystem resides in a Mott insulating state; consequently, the correlated Fermi surface (FS) topology is strongly renormalized, leaving the low-energy physics dominated by the $\mathrm{Ni-}e_g$ orbitals of the BL subsystem. We then construct a quasiparticle (QP) model to implement our symmetry-based approach. Our results reveal that two-gap superconductivity occurs within the BL subsystem, reminiscent of pressurized $\mathrm{La}_3\mathrm{Ni}_2\mathrm{O}_7$~\cite{mfpr-vlnz}. Furthermore, the suppressed $T_c$ can be attributed to the diminished contribution of the interlayer pairing to this two-gap state, tracking the reduction of the hopping ratio $|t_{\perp}^z/t_{\parallel}^x|$, where $t_{\parallel}^x$ is the intralayer hopping between the Ni-$3d_{x^2-y^2}$ orbitals.

\begin{figure}
 \centering{}\includegraphics[width=1\columnwidth,totalheight=1.0\columnwidth,keepaspectratio]{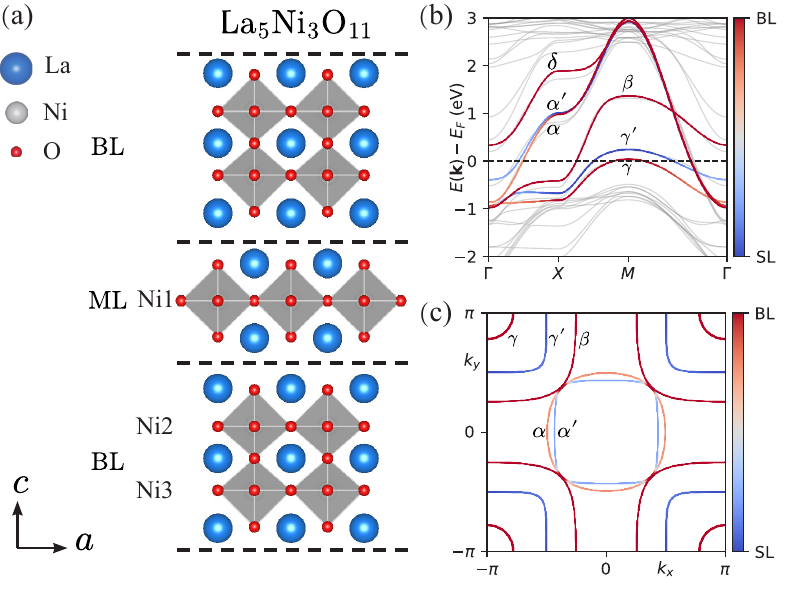}
  \caption{(a) Crystal structure of La$_5$Ni$_{3}$O$_{11}$ in the $P4/mmm$ space group drawn using VESTA~\cite{Momma:db5098}. The crystal structure of La$_{5}$Ni$_{3}$O$_{11}$ is adopted from Ref.~\cite{shiPressureInducedSuperconductivity2025}, corresponding to a pressure of $21$~GPa. (b)–(c) Layer-resolved BS and FS from DFT.\label{fig:1}} 
\end{figure}

\emph{Correlated electronic structure.}---%
Since the electronic structure of ML-BL $\mathrm{La}_5\mathrm{Ni}_3\mathrm{O}_{11}$ has recently been shown to be strongly renormalized by electronic correlations~\cite{k83c-qgx1,leonov2026electronicstructurequasiparticlerenormalizations,1412-nfzm,zhangPairingMechanismSuperconductivity2026b}, we begin with a lattice model incorporating strong on-site repulsion,
$H = H_{\mathrm{kin}} + H_{\mathrm{rep}} $,
where $H_{\mathrm{kin}}$ and $H_{\mathrm{rep}}$ denote the kinetic energy and on-site repulsion, respectively.  In general, the non-interacting kinetic term can be written as $H_{\mathrm{kin}} = \sum_{\boldsymbol{k}s}\psi_{\boldsymbol{k}s}^\dagger\left(\mathcal{H}_{\boldsymbol{k}}-\mu I\right)\psi_{\boldsymbol{k}s}$. Here the basis $\psi_{\boldsymbol{k}s} = (c_{1z,s},c_{1x,s},c_{2z,s},c_{2x,s},c_{3z,s},c_{3x,s})^{T}$, with $z$ and $x$ respectively denoting the $d_{z^2}$ and $d_{x^2-y^2}$ orbitals. The subscripts $s$ denote the spin and $1,2,3$ respectively represent the Ni1, Ni2 and Ni3 atoms, as depicted in Fig.~\ref{fig:1}. To capture the electron-correlation effects, we consider a Slater-Kanamori-type interaction $H_{\mathrm{rep}}$~\cite{lechermann_electronic_2023}, where the constrained random phase approximation (cRPA) method~\cite{KURITA2023108854, NAKAMURA2021107781} is employed, yielding $(U,U',J_{H},J')=(4.8, 3.3, 0.6, 0.6)$~eV for Ni1 and $(4.0, 2.7, 0.5, 0.5)$~eV for Ni2(3).
Thus, each inequivalent Ni site can be treated as a single-site impurity problem, allowing us to apply a charge self-consistent DFT+DMFT approach to reveal the renormalized band structures (BSs). 

 On the one hand, the non-interacting six-orbital TB model is obtianed by Wannier downfolding of the DFT dispersions~\cite{MOSTOFI20142309, Pizzi_2020}. Up to the next-nearest neighbors, the resulting TB BS and FS are shown in Figs.~\ref{fig:1}(b) and (c). Specifically, five Fermi pockets emerge, originating from the Ni-$e_{g}$ orbitals in both the ML and BL subsystems. The $\gamma$ and $\gamma'$ pockets are dominated by the $d_{z^2}$ orbitals, whereas the $\alpha$, $\beta$ and $\alpha'$ pockets are dominated by the $d_{x^2-y^2}$ orbitals~\cite{feng2026supplemental}.

\begin{figure}
 \centering{}\includegraphics[width=1\columnwidth,totalheight=2.0\columnwidth,keepaspectratio]{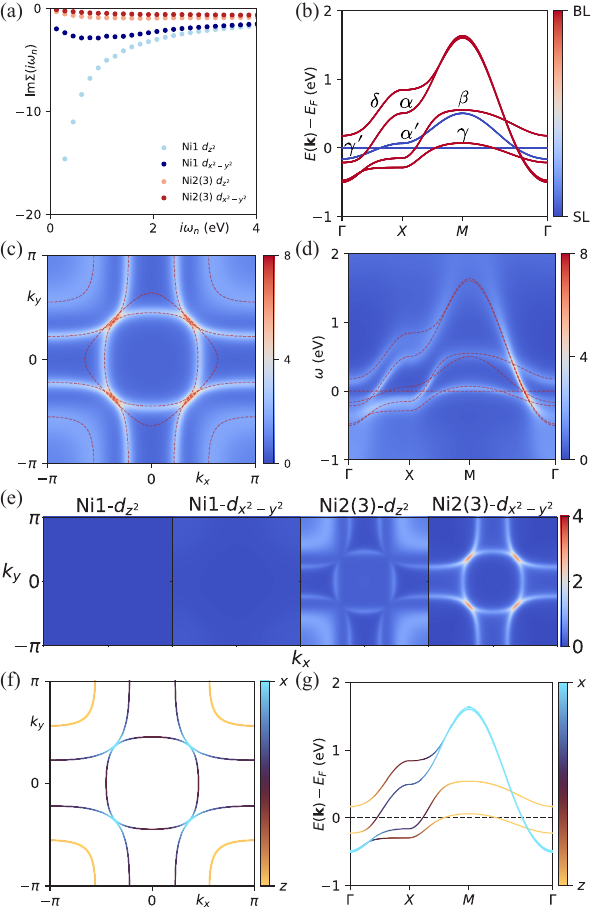}
  \caption{(a) Imaginary part of the Matsubara self-energies of the site-resolved local Ni-$e_g$ orbitals. (b) Contribution of the ML and BL systems to the QP BS after renormalization. (c) Comparison between the QP FS and the spectral function $A(\boldsymbol{k},0)$. (d) Comparison between the QP BS and $A(\boldsymbol{k},w)$ for the whole system. (e) Orbital-resolved correlated FS $A_{\sigma,\sigma}(\boldsymbol{k},0)$. (f) Orbital contribution to the QP FS and (g) BS of the BL subsystem. The filling $n=1.46$ per Ni-$e_g$ orbitals demonstrates that the BL subsystem is hole-doped relative to the undoped filling of $n=1.5$.
  \label{fig:2}}
\end{figure}

On the other hand, DMFT calculations at $T=150~\mathrm{K}$ reveal a clear orbital-selective behavior in $\mathrm{La}_5\mathrm{Ni}_3\mathrm{O}_{11}$. The imaginary part of the self-energy $\mathrm{Im}\Sigma(i\omega_n)$ for the $\mathrm{Ni1-}3d_{z^2}$ orbitals exhibits a sharp divergence in the low-frequency limit, as shown in Fig.~\ref{fig:2}(a), signifying its Mott insulating character. Moreover, $\mathrm{Im}\Sigma(i\omega_n)$ for the Ni1-$3d_{x^2-y^2}$ orbitals remains finite at low Matsubara frequencies, reflecting strong incoherent scattering and reduced QP coherence.
Utilizing the maximum entropy method, the real-frequency self-energy $\Sigma(\omega)$ is obtained. Subsequently, the $\boldsymbol{k}$-resolved spectral function $A(\boldsymbol{k},\omega) = -\frac{1}{\pi} \mathrm{Im}[\mathrm{Tr}\,G(\boldsymbol{k},\omega)]$ with $G(\boldsymbol{k},\omega) = [(\omega+\mu)I-\mathcal{H}(\boldsymbol{k})-\Sigma(\omega)]^{-1}$ exhibits three correlated Fermi pockets, contrasting with the five identified by DFT, as shown in Figs.~\ref{fig:2}(c)-(d). 

To further analyze this orbital-selective behavior, we evaluate the diagonal QP spectral weight matrix $Z \approx \left( I - \left. \partial\mathrm{Re}\Sigma(\omega)/\partial\omega \right|_{\omega\to 0} \right)^{-1}$, which enters the effective QP Hamiltonian~\cite{Yue_2025}:
\begin{equation}
\mathcal{H}^{\mathrm{QP}}(\boldsymbol{k})=Z^{1/2}\left[\mathcal{H}(\boldsymbol{k})-\mu I +\mathrm{Re}\Sigma(0)\right]Z^{1/2}.
\label{eq:1}
\end{equation}
For the ML subsystem, the vanishing weight $Z_{1z,1z}\approx 0$ indicate that the $\mathrm{Ni1-}d_{z^2}$ orbital is fully Mott-insulated. 
Conversely, although $Z_{1x,1x}\approx 0.35$, the $\mathrm{Ni1-}d_{x^2-y^2}$ orbital fails to form a normal metallic state; its pronounced low-frequency imaginary self-energy drives a high scattering rate, resulting in an incoherent metallic state. 
In contrast, the BL subsystem  maintains robust coherent quasi-particle features, with $Z_{2(3)z,2(3)z} \approx 0.32$ and $Z_{2(3)x,2(3)x} \approx 0.50$. The QP BS is shown in Fig.~\ref{fig:2}(b). Moreover, the spectral functions in Fig.~\ref{fig:2}(e) also suggest that the electronic properties of the two subsystems can be analyzed separately. Since the ML subsystem may not host well-defined quasiparticles near the Fermi level that can form Cooper pairs, we therefore focus exclusively on superconductivity arising from the BL subsystem and adopt the approximation $H_{\mathrm{kin}} + H_{\mathrm{rep}} \approx H_{\mathrm{QP}}$.

The effective QP model $\mathcal{H}^{\mathrm{QP}}_{\mathrm{BL}}(\boldsymbol{k})$ for the BL subsystem can then be constructed by separating the  $e_g$ orbitals in Ni2 and Ni3, which yields \cite{feng2026supplemental}:
 \begin{align}
\mathcal{H}^{\mathrm{QP}}_{\mathrm{BL}}({\boldsymbol{k}})&=T_{\boldsymbol{k}}^{z}\Gamma_{0,(0+z),0}/2 \nonumber  +T_{\boldsymbol{k}}^{x}\Gamma_{0,(0-z),0}/2+V_{k}\Gamma_{0,x,0} \nonumber \\
&+T_{\boldsymbol{k}}'^{z}\Gamma_{x,(0+z),0}/2+T_{\boldsymbol{k}}'^{x}\Gamma_{x,(0-z),0}/2+V_{k}'\Gamma_{x,x,0}, \label{eq:2}
 \end{align}
 with
\begin{align}
T_{\boldsymbol{k}}^{x/z} &=\epsilon^{z}+2t_{1,\parallel}^{x/z}\left(\cos k_{x}+\cos k_{y}\right) + 4t_{2,\parallel}^{x/z}\cos k_{x}\cos k_{y}, \nonumber \\
 V_{\boldsymbol{k}}&=2t_{1,\parallel}^{xz}(\cos k_{x}-\cos k_{y}),\nonumber \\
T_{\boldsymbol{k}}'^{x/z} &=t_{1,\perp}^{x/z}+2t_{2,\perp}^{x/z}\left(\cos k_{x}+\cos k_{y}\right)+4t_{3,\perp}^{x/z}\cos k_{x}\cos k_{y}, \nonumber \\
V_{\boldsymbol{k}}' &=2t_{1,\perp}^{xz}\left(\cos k_{x}-\cos k_{y}\right).  \nonumber
 \end{align}
where we have used $\Gamma_{i,j,k}=\rho_{i}\sigma_{j}s_{k}$ as a compact form, with $\rho_{i}$, $\sigma_{j}$, and $s_{k}$ ($i,j,k=0,x,y,z$) representing the Pauli matrices acting on the layer, orbital, and spin degrees of freedom, respectively. The subscripts $\parallel$ and $\perp$ refer to intra-layer and inter-layer hoppings, respectively, while the superscripts $z$, $x$, and $xz$ denote hoppings within the $d_{x^{2}-y^{2}}$ orbitals, the $d_{z^2}$ orbitals, and orbital hybridization, respectively. The basis operator here is defined as $\psi_{\boldsymbol{k}}^{\dagger} = \left[c_{2z\uparrow}^{\dagger},\, c_{2z\downarrow}^{\dagger},\, c_{2x\uparrow}^{\dagger},\, c_{2x\downarrow}^{\dagger},\, c_{3z\uparrow}^{\dagger},\, c_{3z\downarrow}^{\dagger},\, c_{3x\uparrow}^{\dagger},\, c_{3x\downarrow}^{\dagger}\right]$. This QP model shares the same matrix structure as the original TB model but utilizes a distinct set of parameter values, as summarized in Table~\ref{tab:1}. Notably, after renormalization, the interlayer hopping between the Ni2-$d_{z^2}$ orbitals is no longer the largest hopping amplitude, in stark contrast to pressurized La$_3$Ni$_2$O$_7$. 

\begin{table}
\renewcommand{\arraystretch}{1.5}
\caption{Parameter values of the original TB and renormalized QP models for the BL subsystem in La$_5$Ni$_3$O$_{11}$. \label{tab:1}}
\begin{ruledtabular}
\begin{tabular}{cccccccc}

 & $\ensuremath{\epsilon^{z}}$ & $\ensuremath{\epsilon^{x}}$ & $\ensuremath{t_{1,\parallel}^{x}}$ & $\ensuremath{t_{2,\parallel}^{x}}$ & $\ensuremath{t_{1,\parallel}^{z}}$ & $\ensuremath{t_{2,\parallel}^{z}}$ & $\ensuremath{t_{1,\parallel}^{xz}}$\tabularnewline
\hline 
TB & $0.317$ & $0.712$ & $-0.487$ & $0.071$ & $-0.113$ & $-0.017$ & $0.240$\tabularnewline
\hline 
QP & $0.116$ & $0.385$ & $-0.263$ & $0.038$ & $-0.041$ & $-0.006$ & $-0.107$\tabularnewline
\hline 
 & $\ensuremath{t_{1,\perp}^{x}}$ & $\ensuremath{t_{2,\perp}^{x}}$ & $\ensuremath{t_{3,\perp}^{x}}$ & $\ensuremath{t_{1,\perp}^{z}}$ & $\ensuremath{t_{2,\perp}^{z}}$ & $t_{3,\perp}^{z}$ & $\ensuremath{t_{1,\perp}^{xz}}$\tabularnewline
\hline 
TB & $0.016$ & $-0.001$ & $0.002$ & $-0.631$ & $0.015$ & $0.008$ & $-0.035$\tabularnewline
\hline 
QP  & $0.009$ & $-0.000$ & $0.001$ & $-0.230$ & $0.006$ & $0.003$ & $-0.016$\tabularnewline

\end{tabular}
\end{ruledtabular}
\end{table} 
 
\emph{Symmetry-allowed superconducting pairings.}---%
Here, we adopt a phenomenological symmetry-based approach to investigate the pairing symmetry in La$_5$Ni$_3$O$_{10}$. Within this framework, superconducting pairing is driven by effective short-range attractive interactions, while the gap structure is constrained by the experimentally measured $T_c$ and the crystal symmetry~\cite{can_high_temperature_2021,mfpr-vlnz}. We start from a multi-orbital antiferromagnetic exchange interaction $ H_J = J\sum_{\langle ij\rangle \alpha\beta} \left( S_{i\alpha}\cdot S_{j\beta} -\frac{1}{4}n_{i\alpha}n_{j\beta} \right)$. In momentum space, the singlet pairing channel of this interaction can be expressed as a short-range attractive interaction~\cite{feng2026supplemental}. To explore possible odd-parity and triplet pairing instabilities, we further generalize it to an effective short-range interaction with explicit orbital and spin dependence (see Supplementary Notes for details~\cite{feng2026supplemental}),
\begin{align}
 & H_{\mathrm{attr}}=
  \sum_{\boldsymbol{kk}'\alpha\beta ss'\boldsymbol{\delta_r}} V_{\mathrm{eff}} c_{\boldsymbol{k}\alpha s}^{\dagger}c_{-\boldsymbol{k}\beta s'}^{\dagger}c_{-\boldsymbol{k}'\beta s'}c_{\boldsymbol{k}'\alpha s}, \label{eq:3}
\end{align}
where $V_{\mathrm{eff}}=\frac{-V_{\alpha\beta ss'}}{N}e^{-i(\boldsymbol{k}-\boldsymbol{k}')\cdot\boldsymbol{\delta_r}}$ is the factorized interaction strength, with $\alpha$ ($\beta$) denoting the orbital degrees of freedom. Here, $N$ is the number of discrete momentum $\boldsymbol{k}$ (or $\boldsymbol{k}'$);  $\boldsymbol{\delta_r}$ denotes the relative displacement vector connecting the two sites involved in the short-range pairing interaction. Using the Hubbard–Stratonovich decoupling at the mean-field level, the BdG Hamiltonian is subsequently given by
$\mathcal{H}_{\boldsymbol{k}}^{\mathrm{BdG}} = \mathcal{H}_0 + \mathcal{H}_\Delta$ in the Nambu basis $\Psi^\dagger_{\mathbf{k}}=\left[\psi^\dagger_{\mathbf{k}},\psi_{-\mathbf{k}}\right]$,
with
$\mathcal{H}_0 = \mathrm{diag}\{\mathcal{H}_{\mathrm{BL}}^{\mathrm{QP}}(\boldsymbol{k}), -\mathcal{H}_{\mathrm{BL}}^{\mathrm{QP}}(-\boldsymbol{k})^{*}\}$
and
$\mathcal{H}_\Delta = \mathrm{offdiag}\{\tilde{\Delta}(\boldsymbol{k}), \tilde{\Delta}^\dagger(\boldsymbol{k})\}$. 
From a symmetry perspective, the pairing matrix $\tilde{\Delta}$ is constrained by crystal symmetry. We therefore introduce a subscript $i$ to label the independent symmetry-allowed pairing types, i.e.,
$\tilde{\Delta}_{i}=\Delta_{i}\gamma(\mathbf{k})\rho_{l}\sigma_{m}s_{n}$, whose pairing symmetry is determined by the pairing harmonic $\gamma(k)$.  
Following the path integral formulation, the corresponding free energy is then determined by~\cite{feng2026supplemental}
 \begin{equation}
 \mathcal{F}_{\mathrm{S}}(\Delta_i)=\frac{Nn}{V_{i}}\left|\Delta_i\right|^{2}-2k_{B}T\sum_{\boldsymbol{k}\alpha}\ln\left[2\cosh(\beta E_{\boldsymbol{k}\alpha}^{\mathrm{BdG}}/2)\right],\label{eq:6}
 \end{equation}
 where $E_{\boldsymbol{k}\alpha}^{\mathrm{BdG}}$ is the eigenvalues of the BdG Hamiltonian; $\beta=1/k_{B}T$ is the inverse temperature; and $n$ is the number of nonzero elements in $\tilde{\Delta}_{i}$. When $\Delta_{i}=0$, Eq.~\eqref{eq:6} reduces to the non-superconducting free energy reference $\mathcal{F}_{\mathrm{N}}$.  The equilibrium gap value $\Delta_{i}$ can be identified by minimizing $\mathcal{F}_{\mathrm{S}}(\Delta_{i})$, which leads to the self-consistent BCS gap equation $\Delta_i=-\frac{V_{i}}{Nn}\sum_{\boldsymbol{k}}\mathrm{Tr}\left[\frac{\partial\mathcal{H}_{\boldsymbol{k}}^{\mathrm{BdG}}}{\partial\Delta_{i}}U_{\boldsymbol{k}}n_{F}(E_{\boldsymbol{k}}^{\mathrm{BdG}})U_{\boldsymbol{k}}^{\dagger}\right]$,
 where $n_F$ is the diagonal Fermi distribution matrix for $E_{\boldsymbol{k}}^{\mathrm{BdG}}$. Here $U_{\boldsymbol{k}}$ is the unitary matrix that diagonalizes the BdG Hamiltonian. The favored pairing symmetry is determined by comparing the BCS condensation energy of each independent pairing type, $\mathcal{F}_{\mathrm{BdG},i} = \min\big(\mathcal{F}_{\mathrm{S}}(\Delta_{i})\big) - \mathcal{F}_{\mathrm{N}}$. 
 
Specifically, the symmetry-allowed matrix representations of $\tilde{\Delta}_{i}$ are obtained using group-theoretical analysis, subject to the constraints imposed by the $p4/mmm$ layer-group symmetry, time-reversal symmetry, and particle-hole symmetry. All of the symmetry-allowed matrix representations of $\tilde{\Delta}_{i}$ are listed in the Supplementary Notes~\cite{feng2026supplemental}. The interaction strength $V_i$ for each $\tilde{\Delta}_{i}$ is determined by minimizing the BCS gap equation to reproduce the experimental $T_c$. However, not all resulting $V_i$ values are physically reasonable. For singlet pairings, $V_i$ originates from the antiferromagnetic superexchange interaction of the $t$-$J$ model and is therefore bounded by the characteristic exchange scale $J \sim 4|t|^2/U$. For triplet pairings, the effective attractive interaction does not arise directly from the antiferromagnetic superexchange channel; nevertheless, we impose the same energy scale as a phenomenological upper bound, motivated by the expectation that triplet pairing interactions in multiorbital nickelates are generally weaker than the dominant singlet superexchange interaction. By further evaluating $\mathcal{F}_{\mathrm{BdG},i} $ for all feasible pairing types, we identify the leading one. This dominant type, together with other compatible feasible candidates, forms the final superconducting pairing configuration.

 \begin{table}
 	\renewcommand{\arraystretch}{1.5}
	\caption{Feasible symmetry-allowed superconducting pairing types in different channels (CH). The Pauli matrices are defined as $\Gamma_{l,i,j,k}=\tau_{l}\rho_{i}\sigma_{j}s_{k}$, where $\tau_{l}$ denotes the particle-hole degree of freedom. Values of $V$ are solved from the BCS equation to reproduce $T_c\approx 64$~K, with $J\approx4|t|^2/U$ serving as the feasibility threshold. The infeasible $\Delta_{d_{xy}}$ pairing in the $B_{2g}$ channel ($V>J$) is retained for completeness to illustrate our exclusion mechanism.\label{tab:2}}
	\centering{}%
	\begin{ruledtabular}
	\begin{tabular}{cccc}
		
		 CH& $\mathcal{H}_{\Delta}$ & $V$(eV) & $J$(eV)\tabularnewline
		\hline 
		${A_{1g}}$ & $\Delta_{s_{z^2}}\Gamma_{y,x,(0+z)/2,y}$ & $0.212$ & $0.379$\tabularnewline 
		${A_{1g}}$ & $\Delta_{s_{x^2+y^2}}\left[\left(\cos k_{x}+\cos k_{y}\right)\Gamma_{y,0,(0-z),y}\right]$ & $0.181$ &$0.226$\tabularnewline
		${B_{1g}}$ & $\Delta_{d_{x^2-y^2}}\left[\left(\cos k_{x}-\cos k_{y}\right)\Gamma_{y,0,(0-z),y}\right]$ & $0.148$ &$0.226$\tabularnewline
		${B_{2g}}$ & $2\Delta_{d_{xy}}\sin k_{x}\sin k_{y} \Gamma_{y,0,(0-z),y}$ & $0.094$ &$0.055$ \tabularnewline
		${A_{1u}}$ & $\Delta_{p}\left[\sin k_{x}\Gamma_{x,0,(0-z),z}+\sin k_{y}\Gamma_{y,0,(0-z),0}\right]$ & $0.113$ &$0.226$ \tabularnewline
		${A_{2u}}$ & $\Delta_{p}\left[\sin k_{x}\Gamma_{y,0,(0-z),0}-\sin k_{y}\Gamma_{x,0,(0-z),z}\right]$ &$0.113$ &$0.226$ \tabularnewline
		${B_{1u}}$ & $\Delta_{p}\left[\sin k_{x}\Gamma_{x,0,(0-z),z}-\sin k_{y}\Gamma_{y,0,(0-z),0}\right]$  &$0.113$ &$0.226$ \tabularnewline
		${B_{2u}}$ & $\Delta_{p}\left[\sin k_{x}\Gamma_{y,0,(0-z),0}+\sin k_{y}\Gamma_{x,0,(0-z),z}\right]$ &$0.113$ &$0.226$ \tabularnewline
		 
	\end{tabular}
	\end{ruledtabular}
\end{table}

\begin{figure}
\includegraphics[width=1\columnwidth,keepaspectratio]{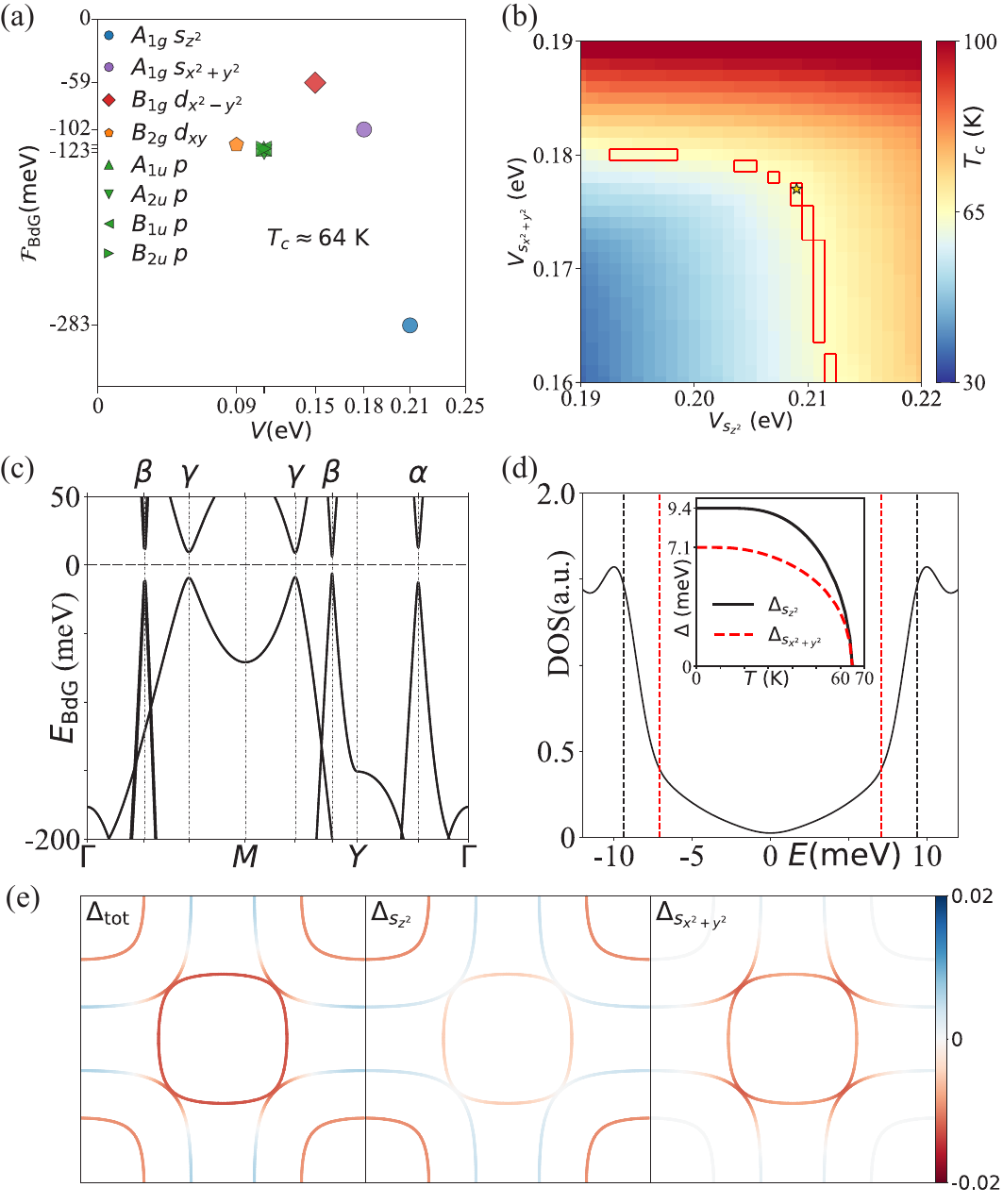}
  \caption{(a) Minimum BCS condensation energy of symmetry-allowed pairing types at $T_c \approx 64$~K. 
   (b) Superconducting transition temperature $T_c$ as a function of the two-gap pairing parameters. The red rectangle marks the parameter region corresponding to $T_c \approx 64$ K. The yellow star indicates $\{V_{s_{z^2}}, V_{s_{x^2+y^2}}\} = \{0.209, 0.177\},\text{eV}$, where the BCS condensation energy is minimized for the two-gap superconducting state.
	(c) Superconducting BS calculated at the marked parameters, with $\{\Delta_{s_{z^2}}, \Delta_{s_{x^2+y^2}}\} = \{9.4, 7.1\}$~meV, yielding gap sizes of $\sim 12.1$ and $9.4$~meV at the $\beta$ and $\gamma$ pockets along $\Gamma$-$M$, respectively.
  	(d) DOS of the superconducting state. 
  	(e) Projected gap functions $\tilde{\Delta}_{\mathrm{tot}} = \tilde{\Delta}_{s_{z^2}} + \tilde{\Delta}_{s_{x^2+y^2}}$ on the QP FS.
  \label{fig:3}}
\end{figure}

Specifically, up to the next-nearest neighbors, the feasible symmetry-allowed pairing types are listed in Table~\ref{tab:2}. The corresponding $\mathcal{F}_{\mathrm{BdG}} $ are depicted in Fig.~\ref{fig:3}(a), which identifies the $s_{z^2}$-wave pairing ($\Delta_{s_{z^2}}$) in the $A_{1g}$ channel as the leading instability, the same as pressurized La$_3$Ni$_2$O$_7$ in our previous study~\cite{mfpr-vlnz}. Since the inclusion of the $s_{x^2+y^2}$-wave pairing component ($\Delta_{s_{x^2+y^2}}$) further lowers the BCS condensation energy relative to the single-gap state, it is energetically favored to coexist with the leading pairing type. For the resulting two-gap superconducting state, we find $\mathcal{F}_{\mathrm{BdG}}\approx -345$~meV at $T_c\approx64$~K as depicted by the marked point in Fig.~\ref{fig:3}(b), substantially lower than  $\mathcal{F}_{\mathrm{BdG}}\approx -283$~meV for the single-gap state shown in Fig.~\ref{fig:3}(a). The two-gap superconductivity leads to a fully gapped superconducting Bogoliubov spectrum (BS) along the high-symmetry path, as shown in Fig.~\ref{fig:3}(c), consistent with ARPES measurements \cite{doi:10.1126/science.adw8329}, which also show a nodeless gap along this path. Notably, accidental nodal points can still emerge within the Brillouin zone (BZ). To examine this, the gap functions are projected onto the FS via $\tilde{\Delta}_{aa}^{\mathrm{proj}}(\boldsymbol{k}_{F}) = U_{a}^\dagger(\boldsymbol{k}_{F}) \tilde{\Delta}(\boldsymbol{k}_{F}) \mathcal{T}[U_{a}(\boldsymbol{k}_{F})]$, where $U_{a}$ is the eigenvector for $\mathcal{H}$, with the subscript $a=\alpha,\beta,\gamma,\delta$ denoting the four bands~\cite{mfpr-vlnz, yang_possible_2023}. As shown in Fig.~\ref{fig:3}(e), the projected  gap $\tilde{\Delta}_{\mathrm{tot}} $ exhibits an $s_{\pm}$-wave pairing symmetry, with the accidental nodes residing on the $\beta$ pocket. Moreover, the density of states (DOS) exhibits a V-shaped nodal gap in Fig.~\ref{fig:3}(d), in good agreement with the tunneling spectra observed in STM/STS experiments~\cite{doi:10.1126/sciadv.aeg2429}. 

\begin{figure}
 \centering{}\includegraphics[width=1\columnwidth,totalheight=1.0\columnwidth,keepaspectratio]{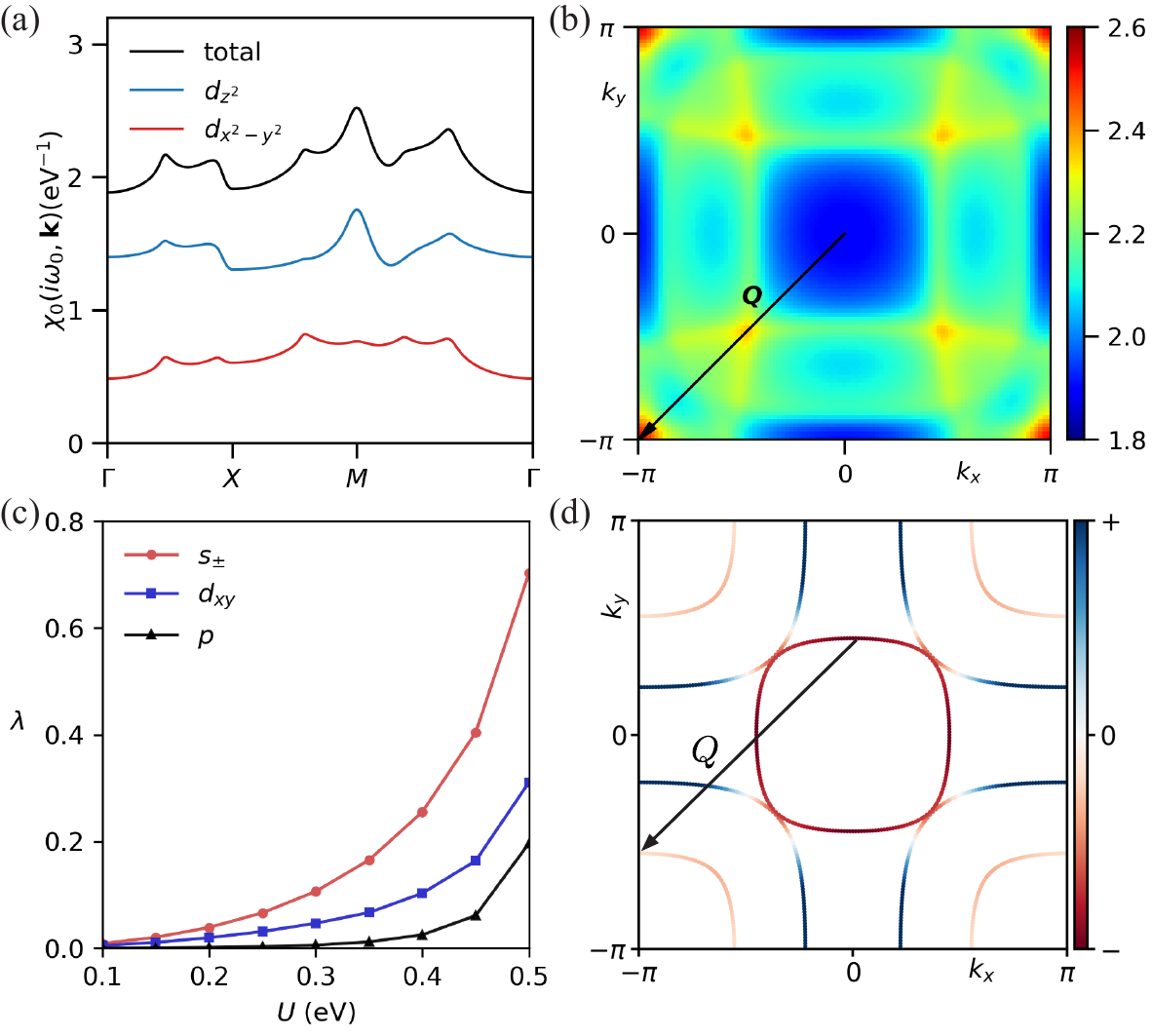}
  \caption{RPA results. (a) Total and orbital-projected bare static susceptibilities $\chi_0(i\omega_0,\boldsymbol{k})$ at $T=116$~K along the high-symmetry path. (b) Total $\chi_0(i\omega_0,\boldsymbol{k})$ in the 2D BZ. (c) The eigenvalues $\lambda$ of the linearized Eliashberg equation calculated with $J_H=J'=0.15U$ and $U'=U-2J_H$. (d) Gap function at $U=0.5$~eV for the leading superconducting instabilities, which exhibit $s_{\pm}$-wave symmetry. 
  \label{fig:4}}
\end{figure}

\emph{Discussion on the possible pairing mechanism.}---%
Our symmetry analysis reveals that the superconducting state of the BL subsystem is dominated by the $\Delta_{s_{z^2}}$ pairing, with the subleading $\Delta_{s_{x^2+y^2}}$ pairing emerging simultaneously. The Pauli matrix representations in Table~\ref{tab:2} explicitly identify $\Delta_{s_{z^2}}$ (i.e., $\Delta_{\perp}^{z}$) as an interlayer pairing between $d_{z^2}$ orbitals, and $\Delta_{s_{x^2+y^2}}$ (i.e., $\Delta_{\parallel}^{x}$) as an intralayer pairing between $d_{x^2-y^2}$ orbitals~\cite{feng2026supplemental}. To discuss the possible pairing mechanism, we first focus on the leading type $\Delta_{s_{z^2}}$.  This pairing type originates from the intraorbital scattering within the $d_{z^2}$ orbitals mediated by spin fluctuations. As shown in Fig.~\ref{fig:3}(e), owing to the orbital composition of the FS, $\Delta_{s_{z^2}}$ is predominantly hosted by the $d_{z^2}$-dominated $\gamma$ pocket, whereas its distribution on the $\alpha$ and $\beta$ pockets is relatively suppressed. Crucially, the orbital hybridization within the $\alpha$ and $\beta$ pockets effectively transforms the intraorbital scattering into interband scattering. Consequently, the $\Delta_{s_{z^2}}$ pairing becomes intimately tied to the nesting between the $\gamma$ and $\beta$($\alpha$) pockets, underscoring the pivotal role of the $\gamma$ pocket in stabilizing the $s_{\pm}$-wave state. This analysis can be verified by the RPA calculations~\cite{Strand2019, PARCOLLET2015398, naturwissenschaftliche_2021}. As shown in Figs.~\ref{fig:4}(a)-(b), the total bare static susceptibility $\chi_{0}$ has a peak at the $M$ point, which means that not only the nesting vector between the $\gamma$ and $\beta$ pockets~\cite{wang_normal_2024}, but also the nesting vector between the $\gamma$ and $\alpha$ pockets can also stabilize the $s_{\pm}$-wave pairing, as shown in Figs.~\ref{fig:4}(c)-(d), similar to the case discussed in Ref.~\cite{xia_sensitive_2025}. This RPA result is also consistent with that obtained in Ref.~\cite{tang2026correlationrenormalizedspinfluctuationpairingstabilization} recently. Notably, the effective interaction $U$ used in the RPA calculation can be viewed as a residual interaction renormalized by a factor of $Z^2$~\cite{Yue_2025}.  Consequently, it is expected to be smaller than the corresponding interaction employed in the DFT+DMFT calculation.

Next, the $\Delta_{s_{x^2+y^2}}$ component serves a subdominant role in the total pairing and does not independently dictate $T_c$. This pairing type is induced by intraorbital scattering within the Ni-$3d_{x^2-y^2}$ orbitals, which are predominantly hosted by the $\alpha$ and $\beta$ pockets. Consequently, this pairing is intimately linked to the interband scattering between the $\alpha$ and $\beta$ pockets. The projected gap pattern on the FS reveals that the corresponding distribution on the $\gamma$ pocket is remarkably suppressed; we thus deduce that the emergence of $\Delta_{s_{x^2+y^2}}$ does not rely on the presence of the $\gamma$ pocket. That is the case in Ref.~{\cite{shao2025pairinggammapocketla3ni2o7film}}.

Finally, we discuss the mechanism underlying the suppressed $T_c$ in $\mathrm{La}_5\mathrm{Ni}_3\mathrm{O}_{11}$ ($64~\mathrm{K}$) and thin-film $\mathrm{La}_3\mathrm{Ni}_2\mathrm{O}_7$ ($40~\mathrm{K}$) relative to pressurized bulk $\mathrm{La}_3\mathrm{Ni}_2\mathrm{O}_7$ ($80~\mathrm{K}$). Within our symmetry-based framework, these three systems exhibit similar pairing configurations but differ in their leading instabilities. While $\mathrm{La}_5\mathrm{Ni}_3\mathrm{O}_{11}$ and pressurized-bulk $\mathrm{La}_3\mathrm{Ni}_2\mathrm{O}_7$ are both dominated by $\Delta_{s_{z^2}}$, thin-film $\mathrm{La}_3\mathrm{Ni}_2\mathrm{O}_7$ favors $\Delta_{s_{x^2+y^2}}$. For thin-film, this mechanism may account for the experimental observation that a smaller in-plane lattice constant leads to a higher $T_c$, whereas a smaller out-of-plane lattice constant does not~\cite{ko_signatures_2025}. For $\mathrm{La}_5\mathrm{Ni}_3\mathrm{O}_{11}$, the ratio $V/J$ listed in Table~\ref{tab:2} for $\Delta_{s_{z^2}}$ ($V_z/J_z \approx 0.56$) is smaller than that for $\Delta_{s_{x^2+y^2}}$ ($V_x/J_x \approx 0.80$). This indicates that achieving a higher $T_c$ via the $\Delta_{s_{x^2+y^2}}$ channel requires $V_x$ to approach the feasibility threshold $J_x$ much more closely than $\Delta_{s_{z^2}}$. Consequently, a dominant $\Delta_{s_{z^2}}$ pairing state inherently promotes a higher $T_c$ because its constraint from $J_z$ is relatively loose, whereas the $\Delta_{s_{x^2+y^2}}$ channel is strictly bounded by this threshold. The suppressed $T_c$ thus reflects the diminished contribution of $\Delta_{s_{z^2}}$ to the overall pairing. At a microscopic level, this shifting balance between the $\Delta_{s_{z^2}}$ and $\Delta_{s_{x^2+y^2}}$ channels may track the electronic hopping ratio $|t_{1,\perp}^{z}/t_{1,\parallel}^{x}|$. For example, the tight-binding parameters reported in Ref.~\cite{PhysRevLett.134.076001} show that the ratio $|t_{1,\perp}^{z}/t_{1,\parallel}^{x}|$ decreases with increasing pressure, consistent with the pressure dependence of $T_c$. Even in the absence of the $\gamma$ pocket, the $\Delta_{s_{x^2+y^2}}$ pairing component can still emerge, although the corresponding $T_c$ may be strongly suppressed. Owing to the two-gap structure, enhancing the interlayer exchange interaction $J_z$ alone may not be sufficient to increase $T_c$. A more effective strategy is to enhance $J_z$ while simultaneously suppressing the intralayer exchange interaction $J_x$. 

To conclude, we have demonstrated a two-gap superconductivity in $\mathrm{La}_5\mathrm{Ni}_3\mathrm{O}_{11}$, and shown that the decrease of $T_c$ compared to pressurized bulk $\mathrm{La}_3\mathrm{Ni}_2\mathrm{O}_7$ is closely related to the decrease of the ratio $|t_{1,\perp}^{z}/t_{1,\parallel}^{x}|$. This conclusion is not only applicable to $\mathrm{La}_5\mathrm{Ni}_3\mathrm{O}_{11}$, but also to superconductivity in BL NiO$_2$ planes for nickelate superconductors. This work highlights the close connection between symmetry and superconductivity, and the application of the present framework to ML and TL nickelates will be explored in future work.

\begin{acknowledgments}
	We thank Dao-Xin Yao, Zhongbo Yan, Cui-Qun Chen and Zhihui Luo for helpful discussions. We also thank the National Supercomputer Center in Guangzhou for supporting this work. This work is supported by Guangdong Basic and Applied Basic Research Foundation (Grant No.~2023A1515110002, 2023A1515011796, 2024A1515011908).
\end{acknowledgments}

\nocite{Merkel2022, QEref1, QEref2, shiPressureInducedSuperconductivity2025, PhysRevLett.77.3865, PhysRevB.88.085117, lechermann_electronic_2023, Seth2016274, PARCOLLET2015398, Held01112007, ZHANG2022108153, ono_symmetry_2019,PhysRevResearch.2.013064,ono_refined_2020, varjas_qsymm_2018,  PhysRevB.106.064509, xia_sensitive_2025}
\bibliography{Feng2026}

\end{document}